\documentclass[twocolumn,nofootinbib]{revtex4-2}
\usepackage{amsmath,amssymb,bm,graphicx,hyperref}
\allowdisplaybreaks[1]

\renewcommand\d{\partial}
\newcommand\+{\dagger}

\newcommand\up{\uparrow}
\newcommand\down{\downarrow}
\newcommand\eps{\varepsilon}
\newcommand\0{{\bm{0}}}
\newcommand\p{{\bm{p}}}
\newcommand\q{{\bm{q}}}
\renewcommand\r{{\bm{r}}}
\newcommand\N{\mathbb{N}}
\newcommand\E{\mathcal{E}}
\renewcommand\L{\mathcal{L}}

\begin{document}

\title{Semisuper Efimov effect induced by resonant pair exchange in mixed dimensions}

\author{Yusuke Nishida}
\affiliation{Department of Physics, Institute of Science Tokyo,
Ookayama, Meguro, Tokyo 152-8551, Japan}

\date{March 2025}

\begin{abstract}
We introduce a new member to the class of semisuper Efimov effects, where an infinite number of bound states emerge with their binding energies obeying the universal scaling law $E_n\sim e^{-2(\pi n/\gamma)^2}$ for sufficiently high excitation $n\in\N$.
Our system consists of a pair of two-component fermions in three dimensions at infinite scattering length, which furthermore interact with a boson confined in two dimensions so as to form a three-body bound state at zero energy.
When another boson is added, the exchange of the resonant pair of fermions between two bosons leads to the semisuper Efimov effect of four such particles with the scaling exponent $\gamma$ determined by the mass ratio of bosons to fermions.
If bosons exist in three dimensions, an infinite number of bound states do not emerge, but some of them may survive for a large mass ratio, making our findings potentially relevant to two-neutron halo nuclei as well as ultracold atoms.
\end{abstract}

\maketitle

\section{Introduction}
The wave nature of particles in quantum mechanics allows them to form a bound state even though their mean separation far exceeds the potential range.
Such loosely bound states are classically forbidden and generally referred to as quantum halos~\cite{Jensen:2004}.
Because their properties can be universal, i.e., independent of the details of the short-range potential, quantum halos have attracted considerable interest across diverse fields in physics ranging from atomic systems~\cite{Chin:2010,Kornilov:2015} to nuclear systems~\cite{Tanihata:2013,Hammer:2017}.

There are special classes of quantum halos, where an infinite number of bound states emerge and their spatial extensions grow exponentially or faster for higher excited states.
Such arbitrarily large quantum halos are classified into a trio of few-body universality classes~\cite{Nishida:2017}
\begin{alignat}{2}
E_n &\sim e^{-2\pi n/\gamma} &\quad& \text{(Efimov)}, \\\label{eq:scaling}
E_n &\sim e^{-2(\pi n/\gamma)^2} &\quad& \text{(semisuper Efimov)}, \\
E_n &\sim e^{-2e^{\pi n/\gamma}} &\quad& \text{(super Efimov)}
\end{alignat}
according to the scaling law of binding energies for sufficiently high excitation $n\in\N$.
Here $\gamma$ is a universal scaling exponent independent of the potential details, and which universality class each system falls into depends on the statistics and dimensionality of particles and the nature of their interaction.
In particular, three bosons in three dimensions at a two-body resonance in the $s$-wave channel exhibit the Efimov effect~\cite{Efimov:1970}, whereas four bosons in two dimensions at a three-body resonance exhibit the semisuper Efimov effect~\cite{Nishida:2017}, but five bosons in one dimension at a four-body resonance exhibit the Efimov effect again~\cite{Nishida:2010}.
On the other hand, the super Efimov effect is exhibited by three fermions in two dimensions at a two-body resonance in the $p$-wave channel~\cite{Nishida:2013}.
Since their discoveries, various extensions have been made, for example, to mass-imbalanced mixtures~\cite{Efimov:1973,Castin:2010,Bazak:2017,Moroz:2014,Horinouchi:2020}, anyons in two dimensions~\cite{Nishida:2008a}, and mixed-dimensional systems~\cite{Nishida:2008b,Nishida:2011}.
Not only have Efimov states been observed experimentally in the systems of ultracold atoms~\cite{Kraemer:2006,Huang:2014,Pires:2014,Tung:2014} and helium atoms~\cite{Kunitski:2015,Voigtsberger:2014}, but their existence has also been subjected to mathematically rigorous proofs~\cite{Barth:2021}.

The purpose of this Letter is to introduce a new member to the class of semisuper Efimov effects.
Our analysis is motivated by an effective field theory developed recently in Ref.~\cite{Hongo:2022} to extract universal properties of two-neutron halo nuclei.
There a two-neutron halo nucleus was described as a loosely bound state of a core nucleus and a pair of neutrons at large scattering length.
We will show that the exchange of such a resonant pair induces a nearly scale invariant attraction between two core nuclei,
\begin{align}
V(\r) \sim \frac1{r^2\ln r},
\end{align}
which is to lead to the semisuper Efimov effect if core nuclei are sufficiently heavy or confined in two dimensions.

\section{Effective field theory}
Our system consists of two-component fermions in three dimensions as well as bosons in arbitrary dimensions $d=1,2,3$.
Such fermions are described by
\begin{align}
\L_\text{3D} &= \sum_{\sigma=\up,\down}
\psi_\sigma^\+\left(i\d_t+\frac{\nabla^2}{2m}\right)\psi_\sigma
- \frac1f\,\Psi^\+\Psi \notag\\
&\quad + \Psi^\+\psi_\up\psi_\down + \psi_\down^\+\psi_\up^\+\Psi,
\end{align}
where $m$ is the mass of fermions and $\Psi\sim\psi_\up\psi_\down$ is the dimer field.
Its kinetic term is absent, corresponding to a vanishing effective range, and $f$ is the coupling constant related to the scattering length $a$ via
\begin{align}
\frac1f - \int\!\frac{d^3\q}{(2\pi)^3}\frac{m}{\q^2} = -\frac{m}{4\pi a}.
\end{align}
A pair of fermions is assumed to be at infinite scattering length and furthermore interact with a boson so as to form a three-body bound state at zero energy.
Such an interaction is described by
\begin{align}\label{eq:action}
\L_\text{$d$D} &= \phi^\+\left(i\d_t+\frac{\nabla^2}{2M}\right)\phi
+ \Phi^\+\left(i\d_t+\frac{\nabla^2}{2M+4m}-\E_0\right)\Phi \notag\\
&\quad + g\Phi^\+\phi\Psi + g\Psi^\+\phi^\+\Phi,
\end{align}
where $M$ is the mass of bosons and $\Phi\sim\phi\Psi$ is the trimer field consisting of a boson and a dimer~\cite{Hongo:2022}.%
\footnote{Such a pointlike trimer emerges as a consequence of the logarithmic divergence of the normalization integral of the wave function at the origin.
This speciality is actually common to the $s$-wave resonance in four dimensions~\cite{Nussinov:2006,Nishida:2006,Nishida:2007}, the $p$-wave resonance in two dimensions~\cite{Nishida:2008a,Nishida:2013,Moroz:2014}, and the three-body resonance in two dimensions~\cite{Nishida:2017}.}
Its bare energy $\E_0$ has to be tuned so that the trimer has zero binding energy for a given coupling constant $g$ [see Eq.~(\ref{eq:self-energy}) below].
We note that $\hbar=1$ and two-component fermions can be replaced with spinless bosons described by
\begin{align}
\L'_\text{3D} &= \psi^\+\left(i\d_t+\frac{\nabla^2}{2m}\right)\psi
- \frac1f\,\Psi^\+\Psi \notag\\
&\quad + \frac1{\sqrt2}\Psi^\+\psi\psi + \frac1{\sqrt2}\psi^\+\psi^\+\Psi
\end{align}
without changing any results presented in this Letter.

The total action is then provided by $S=\int\!dtd^d\r d^{3-d}\r_\perp[\L_\text{3D}(t,\r,\r_\perp)+\L_\text{$d$D}(t,\r)\delta^{3-d}(\r_\perp)]$, with $\r$ and $\r_\perp$ the $d$-dimensional and $(3-d)$-dimensional coordinates perpendicular to each other, respectively.
This is the effective field theory that captures universal properties of our system at low energy and long wavelength.
In order to develop an intuitive understanding of how the semisuper Efimov effect emerges, we first employ the Born-Oppenheimer approximation assuming $M\gg m$.

\section{Born-Oppenheimer analysis}
\begin{figure}[t]
\includegraphics[width=\columnwidth,clip]{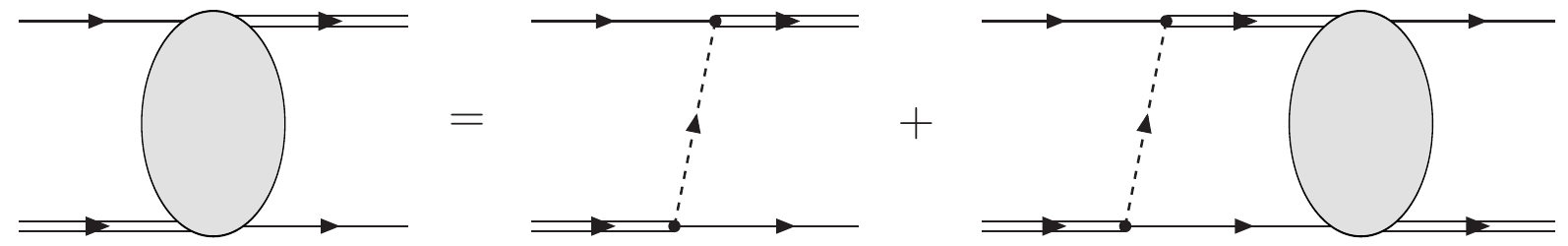}
\caption{\label{fig:scattering}
Feynman diagrams for the scattering $T$ matrix represented by the blob.
Solid, dashed, and double lines represent the propagators of the boson, dimer, and trimer, respectively.}
\end{figure}

The scattering between two bosons is induced by exchanging the resonant pair of fermions as depicted by the Feynman diagrams in Fig.~\ref{fig:scattering}.
Here the on-shell $T$ matrix satisfies the integral equation
\begin{align}\label{eq:integral}
T(E;\p,\p') &= g^2D(E-\eps_\p-\eps_{\p'},-\p-\p') \notag\\
&\quad + \int\!\frac{d^d\q}{(2\pi)^d}g^2D(E-\eps_\p-\eps_\q,-\p-\q) \notag\\
&\qquad \times G_0(E-\eps_\q,-\q)T(E;\q,\p'),
\end{align}
where $E$ is a collision energy in the center-of-mass frame and $\p$ ($\p'$) is an incoming (outgoing) $d$-dimensional momentum of the boson with its energy $\eps_\p=\p^2/2M$.
Furthermore,
\begin{align}
D(p_0,\p) = -\frac{4\pi}{m}\int\!\frac{d^{3-d}\p_\perp}{(2\pi)^{3-d}}
\frac1{\sqrt{\frac{\p^2+\p_\perp^2}{4}-mp_0-i0^+}}
\end{align}
is the propagator of the dimer at infinite scattering length projected onto the $d$-dimensional plane and
\begin{align}
G_0(p_0,\p) = \frac1{p_0-\frac{\p^2}{2M+4m}+i0^+}
\end{align}
is the bare propagator of the trimer for $\E_0=0$.
Its renormalization is to be considered later.

In the limit of $M/m\gg1$ with $E=k^2/M$, the energy dependence of $D(E-\eps_\p-\eps_\q,-\p-\q)$ can be neglected, so that Eq.~(\ref{eq:integral}) is reduced to the Lippmann-Schwinger equation.
The corresponding Schr\"odinger equation reads
\begin{align}
\left(E+\frac{\nabla^2}{M}\right)\chi(\r) = V(\r)\chi(-\r),
\end{align}
where the effective potential is provided by
\begin{align}\label{eq:potential}
V(\r) = \int\!\frac{d^d\p}{(2\pi)^d}\,e^{i\p\cdot\r}g^2D(0,\p)
= -\frac{4g^2}{\pi mr^2}
\end{align}
and the minus sign in $\chi(-\r)$ reflects the fact that a boson and a trimer swap by exchanging a dimer.
Therefore, the resonant pair exchange apparently induces an inverse square attraction (repulsion) in an even-parity (odd-parity) channel, which is scale invariant and leads to the Efimov effect if attractive~\cite{Fonseca:1979}.
However, this conclusion is not completely correct because renormalization of the trimer field is not taken into account.

\begin{figure}[t]
\includegraphics[width=0.5\columnwidth,clip]{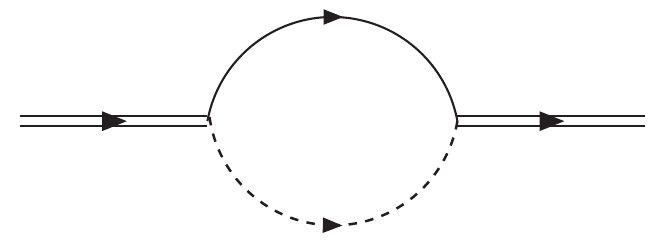}
\caption{\label{fig:self-energy}
Feynman diagram for the trimer self-energy.}
\end{figure}

The trimer self-energy is depicted by the Feynman diagram in Fig.~\ref{fig:self-energy},
\begin{align}\label{eq:self-energy}
\Sigma(p_0,\p) = \int\!\frac{d^d\q}{(2\pi)^d}g^2D(p_0-\eps_\q,\p-\q),
\end{align}
which is quadratically divergent.
Such a divergence can be eliminated by tuning $\E_0$ and a three-body bound state is formed at zero energy when $\E_0+\Sigma(0,\0)=0$.
Consequently, the renormalized propagator of the trimer multiplied by $g^2$ turns out to be
\begin{align}
g^2G(p_0,\p) &= \frac{g^2}{1 + \frac{8g^2}{\pi}\left(\frac{M}{M+2m}\right)^{d/2}
\ln\!\left(\frac{\Lambda}{\sqrt{-m\tilde p_0}}\right)} \notag\\
&\qquad \times \frac1{p_0-\frac{\p^2}{2M+4m}+i0^+},
\end{align}
where a remaining logarithmic divergence is cut off by $\Lambda$ with $\tilde p_0\equiv p_0-\p^2/(2M+4m)+i0^+$.
By comparing the resulting expression to $g^2G_0(p_0,\p)$, the role of renormalization is found to replace the coupling constant with
\begin{align}\label{eq:3-body}
g^2(s) = \frac1{\frac1{g^2} + \frac8\pi\left(\frac{M}{M+2m}\right)^{d/2}s}
\end{align}
at a momentum scale $e^{-s}\Lambda$.
This is the running coupling decreasing logarithmically toward the infrared limit $s\to\infty$.

Such a scale dependence of the coupling necessarily makes the effective potential in Eq.~(\ref{eq:potential}) scale dependent.
By identifying the characteristic scale as $s=\ln(r/r_0)$, with $r_0$ a short-distance cutoff set by the potential range, the Schr\"odinger equation in the $s$-wave channel now reads
\begin{align}\label{eq:schrodinger}
E\,r^{(d-2)/2}\chi(r)
&= \left[-\frac1M\left(\frac{\d^2}{\d r^2} + \frac1r\frac\d{\d r}\right)
+ \frac{(d-2)^2}{4Mr^2} \right. \notag\\
&\quad\left.{} - \frac1{2mr^2\ln(r/r_0)}\right]r^{(d-2)/2}\chi(r)
\end{align}
in the limits of $M\gg m$ and $r\gg r_0$.%
\footnote{If bosons are confined in quasi-low-dimensions, $r_0$ is set by the potential range or the confinement length, whichever is larger.}
Therefore, the resonant pair exchange actually induces a nearly scale invariant attraction that is the inverse square but weakened by the logarithmic correction.
If bosons are confined in $d=2$, the semiclassical quantization condition leads to an infinite number of bound states with their binding energies obeying the scaling law of Eq.~(\ref{eq:scaling}) with the scaling exponent provided by $\gamma=\sqrt{2M/m}$~\cite{Efremov:2014,Efremov:2015,Nishida:2017}.
This is the semisuper Efimov effect of two bosons and two fermions in mixed dimensions.

On the other hand, if bosons exist in $d=3$ or are confined in $d=1$, the nearly scale invariant attraction is hidden behind the inverse square repulsion at $r\to\infty$, so that an infinite number of bound states do not emerge.
However, some of them may survive for $M\gg m$, where the inverse square repulsion is suppressed.
Because the nearly scale invariant attraction is dominant at $\ln(r/r_0)\lesssim2M/m$, bound states with excitation numbers up to $n\sim(2M/m)/\pi$ are expected to remain.
Similarly, if the scattering length $a$ is finite or the trimer has nonzero binding energy $\E$, the nearly scale invariant attraction is to be cut off at long distance by $R_0\sim|a|$ or $1/\sqrt{m|\E|}$, whichever is smaller.
Consequently, only a finite number of bound states up to $n\sim\gamma\sqrt{\ln(R_0/r_0)}/\pi$ are expected to emerge.

\section{Renormalization-group analysis}
Although the Born-Oppenheimer approximation is helpful to obtain an intuitive understanding of how the semisuper Efimov effect emerges in $d=2$, it applies only to the limit of $M\gg m$.
Let us then derive the semisuper Efimov effect from a different perspective with no assumption about the masses of bosons and fermions.

In order for the effective field theory to capture universal properties of our system at low energy and long wavelength, the three-body coupling $g$ between a boson and a dimer is necessary in Eq.~(\ref{eq:action}) because it is marginal in the sense of the renormalization group (RG).
If bosons are confined in $d=2$, there are three more marginal couplings
\begin{align}
\L_\text{2D} = v_2\phi^\+\phi^\+\phi\phi
+ v_4\Phi^\+\phi^\+\phi\Phi + v_6\Phi^\+\Phi^\+\Phi\Phi
\end{align}
that have to be included in our effective field theory.
Here $v_2$, $v_4$, and $v_6$ are the two-body, four-body, and six-body couplings between two bosons, a boson and a trimer, and two trimers, respectively.
We will not consider the six-body coupling further because it is irrelevant to the semisuper Efimov effect of two bosons and two fermions in mixed dimensions.

\begin{figure}[t]
\includegraphics[width=0.9\columnwidth,clip]{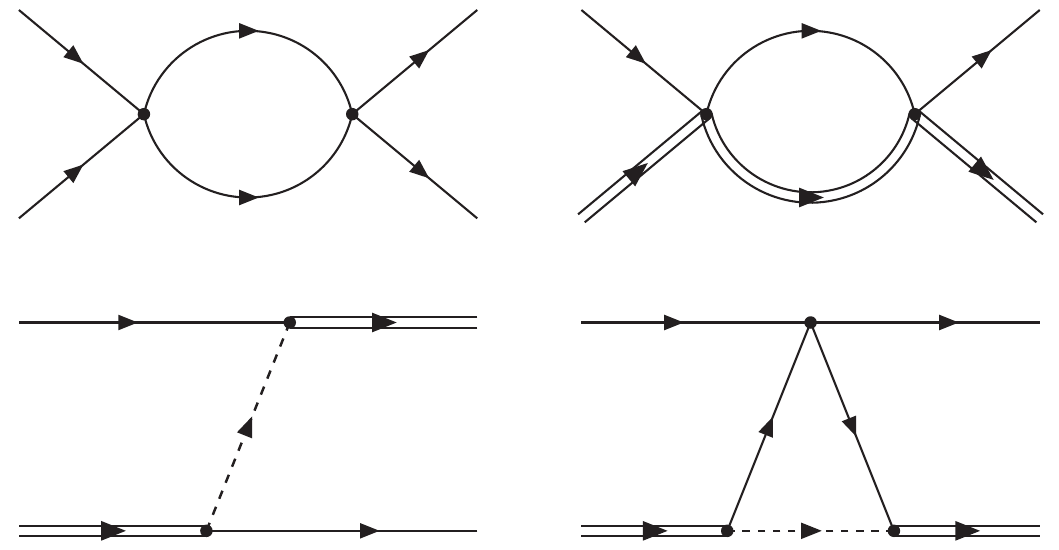}
\caption{\label{fig:renormalization}
Feynman diagrams to renormalize the two-body coupling (top left diagram) and the four-body coupling (the rest of the diagrams).}
\end{figure}

The two-body and four-body couplings are renormalized by the Feynman diagrams depicted in Fig.~\ref{fig:renormalization}.
Consequently, the running of $v_2$ at a momentum scale $e^{-s}\Lambda$ is governed by
\begin{align}
\frac{dv_2}{ds} &= \frac{M}{\pi}v_2^2,
\end{align}
which is solved by
\begin{align}\label{eq:2-body}
v_2(s) = \frac1{\frac1{v_2}-\frac{M}{\pi}s}
\to -\frac\pi{M}\frac1s \quad (s\to\infty).
\end{align}
On the other hand, the running of $v_4$ is governed by
\begin{align}\label{eq:running1}
\frac{dv_4}{ds} &= -\frac8\pi\frac{M}{M+2m}g^2v_4
+ \frac1{2\pi}\frac{M(M+2m)}{M+m}v_4^2 \notag\\
&\quad + \frac8mg^2 + O(g^2v_2),
\end{align}
where the first term arises from the wave-function renormalization of the trimer.
The last term proportional to $g^2v_2$ is not presented explicitly because it is negligible in the infrared limit $s\to\infty$, where the RG equation is reduced to
\begin{align}\label{eq:running2}
\frac{dv_4}{ds} = -\frac{v_4}{s} + \frac1{2\pi}\frac{M(M+2m)}{M+m}v_4^2
+ \pi\frac{M+2m}{Mm}\frac1s + O(s^{-2})
\end{align}
by substituting the asymptotic forms of $g^2$ and $v_2$ for $d=2$ from Eqs.~(\ref{eq:3-body}) and (\ref{eq:2-body}), respectively.

Its solution in the same limit is then provided by
\begin{align}
\sqrt{s}v_4(s) \to -\sqrt{2\pi^2\frac{M+m}{M^2m}}
\cot\!\left(\sqrt{\frac{2(M+2m)^2}{(M+m)m}s}-\theta\right),
\end{align}
where $\theta$ is a nonuniversal constant depending on the details at the ultraviolet scale $s\sim0$.
Therefore, we find that $\sqrt{s}v_4$ is a periodic function of $\sqrt{s}$ and diverges infinitely many times at
\begin{align}
s_n = \frac{(M+m)m}{2(M+2m)^2}(\pi n + \theta)^2
\end{align}
for sufficiently large $n\in\N$.
These divergences of the four-body coupling in its RG flow indicate the existence of characteristic energy scales $E_n\sim e^{-2s_n}\Lambda^2$ in the system of two bosons and two fermions in mixed dimensions.
Such energy scales are naturally identified as their binding energies~\cite{Bedaque:1999a,Bedaque:1999b,Nishida:2013,Moroz:2014,Nishida:2017}, so that the semisuper Efimov effect in Eq.~(\ref{eq:scaling}) is predicted with
\begin{align}\label{eq:exponent}
\gamma = \sqrt{\frac{2(M+2m)^2}{(M+m)m}}.
\end{align}
The resulting scaling exponent is a monotonically increasing function of the mass ratio of bosons to fermions.
In particular, it approaches $\gamma\to\sqrt{2M/m}$ for $M\gg m$, consistent with the Born-Oppenheimer approximation, whereas it is reduced to $\gamma\to2\sqrt2$ in the opposite limit $M\ll m$.%
\footnote{The exact scaling exponent in Eq.~(\ref{eq:exponent}) can also be obtained with the Born-Oppenheimer approximation by keeping $M/m$ finite in Eq.~(\ref{eq:schrodinger}), which is to replace $M$ with $M(M+2m)/(M+m)$ in the reduced mass and $m$ with $Mm/(M+2m)$ in the effective potential.}

Finally, we note that the bosonic $\phi$ field is essential to the semisuper Efimov effect.
If the $\phi$ field were fermionic, the third term on the right-hand side of Eq.~(\ref{eq:running1}) or (\ref{eq:running2}) would acquire a minus sign.
The resulting solution then would turn out to be $\sqrt{s}v_4(s)\to-\sqrt{2\pi^2(M+m)/(M^2m)}$ with no divergences in the infrared limit $s\to\infty$, indicating the absence of an infinite number of bound states.
Similarly, from the perspective of the Born-Oppenheimer approximation, both terms on the right-hand side of Eq.~(\ref{eq:integral}) would acquire minus signs.
The resulting effective potential then would flip its sign and become repulsive in the $s$-wave channel.

\section{Conclusion}
We have introduced a new member to the class of semisuper Efimov effects.
It consists of a pair of two-component fermions (or spinless bosons) in three dimensions at infinite scattering length interacting with each of two bosons confined in two dimensions so as to form a three-body bound state at zero energy.
We showed that the exchange of the resonant pair of fermions between two bosons leads to an infinite number of bound states of four such particles, whose binding energies obey the scaling law of Eq.~(\ref{eq:scaling}) with the scaling exponent determined by the mass ratio of bosons to fermions as in Eq.~(\ref{eq:exponent}).
Although simultaneous fine-tuning of the interactions between two fermions and between fermions and a boson is required, its implementation is not impossible in ultracold-atom experiments with the help of proposed schemes to independently control two-body and three-body interactions~\cite{Daley:2014,Petrov:2014a,Petrov:2014b,Paul:2016}.
Once our system is realized, the emergent semisuper Efimov states may be observed via resonantly enhanced atom losses by detuning the resonant interactions~\cite{Kraemer:2006,Huang:2014,Pires:2014,Tung:2014}.

If bosons exist in three dimensions, an infinite number of bound states do not emerge, but some of them may survive for a large mass ratio.
Such a system is potentially relevant to two-neutron halo nuclei by identifying bosons as core nuclei and fermions as neutrons with a large scattering length $a$ and a small separation energy $-\E$~\cite{Hongo:2022}.
The nearly scale invariant attraction of Eq.~(\ref{eq:schrodinger}) is then induced in a finite range $r_0\ll r\ll|a|,\,1/\sqrt{m|\E|}$ and may serve as an exotic binding mechanism of two core nuclei by exchanging a pair of neutrons.
Our findings in this Letter advance the physics of quantum halos and its universality across atomic and nuclear systems, which are hopefully to stimulate further efforts toward experimental realization and (still lacking~\cite{Barth:2021}) mathematically rigorous proof of the semisuper Efimov effect.

\acknowledgments
This work was supported by JSPS KAKENHI Grant No.~JP21K03384.

\end{document}